\documentclass[11pt,twoside]{article}
\usepackage{graphicx}
\usepackage{epsfig}

\usepackage{hhline}

\def\beq{\begin{equation}}
\def\eeq{\end{equation}}

\def\s{{\,\rm s}}

\def\K{\,{\rm K}}

\def\keV{\,{\rm keV}}

\def\GeV{\,{\rm GeV}}
\def\TeV{\,{\rm TeV}}

\def\cm{\,{\rm cm}}
\def\kms{\,{\rm km/s}}
\def\pb{\,{\rm pb}}

\def\gcm{\,{\rm g/cm^3}}
\def\gc{\,{\rm g/cm^2}}
\def\Gcm{\,{\rm GeV/cm^3}}
\def\nm{\nu_{\mu}}
\def\anm{\bar{\nu}_{\mu}}
\def\m{\bar{m}}

\def\to{\rightarrow}
\newcommand{\lsim}{\lower .7ex\hbox{$\;\stackrel{\textstyle <}{\sim}\;$}}
\newcommand{\gsim}{\lower .7ex\hbox{$\;\stackrel{\textstyle >}{\sim}\;$}}
\def\sv{\left<\sigma v\right>}
\newcommand{\aver}[1]{\left<#1\right>}
\newcommand{\fr}[3]{\left(\frac{#1}{#2}\right)^{\!\! #3}}

\def\N2{$N_2$}
\def\DT{\Delta T}

\date{}

\title{Muon Flux Limits for Majorana Dark Matter Particles}
\author{Konstantin Belotsky$^1$, Maxim Khlopov$^{1,2}$, Chris Kouvaris$^3$\\
\\
{\small 1) Moscow Engineering Physics Institute, Moscow, Russia}\\
{\small Center for Cosmoparticle physics ``Cosmion'', Moscow, Russia}\\
{\small 2) APC laboratory, Paris, France} \\
{\small 3) The Niels Bohr Institute, Copenhagen, Denmark} }

\begin{document}

\maketitle

\begin{abstract}
We analyze the effects of capture of dark matter
 (DM) particles, with successive annihilations, predicted in the minimal walking technicolor model (MWT) by the Sun and the Earth. We show that the Super-Kamiokande (SK)
 upper limit on excessive muon flux disfavors the mass
 interval between 100-200 GeV for MWT DM with a suppressed Standard Model interaction (due to a mixing angle), and the mass interval between 0-1500 GeV for MWT DM
 without such suppression, upon making the standard assumption about the value of the local DM distribution.
 In the first case,
the exclusion interval is found to be very sensitive to the DM distribution parameters and can vanish at the extreme of the acceptable values.
\end{abstract}

\section{\label{introduction}Introduction}

The possibility of breaking the electroweak symmetry in a natural dynamical way in the context of technicolor has been a very appealing one since it was first introduced in the 70s~\cite{Weinberg:1979bn,Susskind:1979up}. It was soon enough realized that in order to overcome the problems of the first models, and in particular the necessity to give mass even to the heaviest Standard Model particles like the top quark, a walking coupling is required. However such a quasi-conformal behavior was associated with a large number of extra flavors coupled to the electroweak sector, making impossible to evade the strict constraints from the electroweak precision measurements. Recently in a series of papers~\cite{Sannino:2004qp,Hong:2004td,Dietrich:2005jn,Dietrich:2006cm}, it has been demonstrated that the above problem can be avoided as soon as the techniquarks transform under higher dimensional representations of the gauge group. More specifically, theories with fermions in the 2-index symmetric representation of the gauge group can accommodate such a quasi-conformal behavior with a very small number of flavors, in contrast to the case where the fermions transform under the fundamental representation. This set of theories does not violate the experimental constraints, thus being an eligible candidate for the upcoming search of the LHC collider.

Since in principle such theories are strongly coupled, a perturbative treatment can offer very little. Non-perturbative techniques and tools should be implemented for the exploration and study of these theories. In this context, the first lattice simulations have been developed~\cite{Catterall:2007yx,DelDebbio:2008wb,DelDebbio:2008zf}. Similarly, low energy effective theories, valid at the scale where LHC will operate, can offer an alternative approach to the problem, and distinct signatures that can rule in or out walking technicolor ~\cite{Gudnason:2006ug,Foadi:2007ue,Belyaev:2008yj}. In addition, holographic methods inspired by the AdS/CFT correspondence can reduce the parameter space of the arbitrary couplings of the effective theories and make more definite phenomenological predictions~\cite{Dietrich:2008ni,Dietrich:2008up}. This last idea is quite promising, especially if one takes into consideration the fact that quasi-conformal theories resemble the exact conformal $N=4$ theory more than QCD, where this method has given results close to the known experimental values.

The simplest of the WTC models, is the one with only two flavors in the techicolor sector, i.e. the techniquarks $U$ and $D$, where they transform under the 2-index symmetric representation of an $SU(2)$ gauge group\footnote{The 2-index representation of the $SU(2)$ is the adjoint one.}. There is an extra family of heavy leptons, i.e. $\nu'$ and $\zeta$ that couple to the electroweak sector, in order to cancel the Witten global anomaly~\cite{Gudnason:2006ug}. Such a theory, although simple in terms of particle content, has a very rich structure. The techniquarks  possess an enhanced $SU(4)$ global symmetry, that includes $SU(2)_L \times SU(2)_R$ as a subgroup. This is due to the fact that the adjoint representation is real. After the chiral symmetry breaking, the vacuum is invariant under an $SO(4)$ symmetry that includes the $SU(2)_V$ as a subgroup. 9 Goldstone bosons emerge from the breaking, three of which are eaten by the $W$ and $Z$ bosons. The remaining 6 bosons come in three pairs of particle-antiparticle, and are $UU$, $UD$, and $DD$, where we have suppressed color and Dirac indices. Their main feature is that although Goldstone bosons, they are not regular mesons, like the pions, since they are composed of two techniquarks, rather than a pair of quark-antiquark. Therefore these Goldstone bosons carry technibaryon number, that can protect the lightest particle from decaying. This fact opens interesting possibilities for dark matter  candidates.

 There is an anomalous free hypercharge assignment for the techniquarks, that makes one of them electrically neutral. For the sake of our study, we choose $D$ to be the one, although the results are identical if we make $U$ instead of $D$ neutral. For the above hypercharge assignment, the corresponding electric charges are $U=1$, $D=0$, $\nu=-1$, and $\zeta=-2$. The first possibility of having a dark matter candidate is the case where the $DD$ (which is electrically neutral) is also the lightest technibaryon~\cite{Gudnason:2006yj}. The technibaryon number of $DD$
can protect it from decaying to Standard Model particles, as long as there are no processes that violate the
technibaryon number below some scale. The sphaleron processes violate the technibaryon number, however, they become ineffective once the temperature of the Universe drops below the electroweak scale. Such a particle can account for the whole dark matter density, as long as its mass is of order TeV. However, such a scenario is excluded by direct dark matter search experiments, since the cross section of $DD$ scattering off nuclei targets is sufficiently large and there should be detected in these experiments. This problem can be avoided in a slightly different version of the MWT~\cite{Ryttov:2008xe}.
Another possibility is to assume that $UU$ is the lightest technibaryon. In this case, it is possible to form electrically bound neutral states between $^4He^{++}$ and $\bar{U}\bar{U}$~\cite{Khlopov:2007ic,Khlopov:2008ty,Blois2008}. Alternatively, it can be  $^4He^{++}$ bound to $\zeta^{--}$. This scenario cannot be excluded by underground detectors and therefore, it is a viable candidate.

In this paper we focus on a yet third possibility. Due to the fact that in WTC the techniquarks transform under the adjoint representation of the gauge group, it is possible to form bound states between a $D$ and technigluons~\cite{CK}. The object $D^aG^a$, where $G$ represents the gluons of the theory, and $a$ runs over the three color states (since it is the adjoint representation), is electrically neutral and colorless. If $D_LG$ has a Majorana mass, a see-saw mechanism is implemented and the mass eigenstates are two Majorana particles, namely a heavy $N_1$ and a light $N_2$. Although the Majorana mass term breaks the technibaryon symmetry, a $Z_2$ symmetry, like the R-parity in neutralinos, protects $N_2$ from decaying. Because $N_2$ is a linear combination of $D_LG$ and $D_RG$ (that does not couple to the electroweak), $N_2$ has a suppressed coupling to the $Z$ boson of the form \beq \frac{\sqrt{g^2+g'^2}}{2}Z_{\mu}\sin^2\theta \bar{N_2}\gamma^5\gamma^{\mu}N_2, \eeq
where $g$ and $g'$ are the electroweak and hypercharge couplings respectively. We have omitted terms that couple
 $N_1$ with $N_2$ as $N_1$ is very heavy and decays to $N_2$ very fast. The mixing angle $\theta$ is defined
  through the relation $\tan 2\theta = 2m_D/M$, where $m_D$ and $M$ are the Dirac and Majorana masses of the $D_LG$ particle. Although the $Z_2$ symmetry can protect $N_2$ from decaying, two $N_2$ can co-annihilate to Standard Model particles through $Z$ boson mediation. For small masses of $N_2$, the main annihilation channel in the early Universe is to pairs of light particles-antiparticles, like quarks and leptons. For larger masses, the dominant channel is annihilation to pairs of $W^+-W^-$. It was shown in~\cite{CK} that because of the $\sin\theta$ dependence of the annihilation cross section, $\sin\theta$ can be chosen accordingly in such a way that the relic density of $N_2$ matches the dark matter density of the Universe. This is depicted in Fig.~\ref{sin}, where $\sin\theta$ is given as function of the mass of $N_2$, in order for $N_2$ to account for the dark matter density. The fact that $N_2$ is a Majorana particle and consequently does not have coherent enhancement in scattering off the nuclei targets in the underground detectors, along with the suppression of the cross section that scales as $\sin^4\theta$, makes $N_2$ evasive from experiments like CDMS for almost any mass of interest. On the same reason this form of dark matter cannot explain the positive results of DAMA/NaI and DAMA/Libra experiments \cite{DAMA}. Under specific conditions, $N_2$ can be excluded as a main dark matter particle only for a small window of masses roughly between 100 and 150 GeV~\cite{Kouvaris:2008hc}. $N_2$ is also susceptible to indirect signatures as those suggested in~\cite{Kouvaris:2007ay}.

In a similar fashion, for a hypercharge assignment where $\nu'$ is neutral, this heavy neutrino can be a dark matter particle if the evolution in the early Universe is dominated by quintessence-like dark energy~\cite{fin}.  However, it was pointed out in~\cite{Kouvaris:2008hc}, that such a candidate even in the case where $\nu'$ is a Majorana particle, is excluded for masses up to 1 TeV (depending on the value of the local dark matter density of the Earth).

We should emphasize here that the investigation and the results of \cite{CK} can also apply for $\nu'$ for the case that it is electrically neutral, if we assume that the left handed $\nu'$ has a Majorana mass and a Dirac mass. Due to the fact that $DG$ (for the hypercharge value that makes it neutral) interacts with the electroweak sector as $\nu'$ does (for a different hypercharge, that makes $\nu'$ neutral), $N_2$ can equally represent the lightest Majorana particle made of $DG$ or $\nu'$.

We would like to stress a few points regarding the models we use in this paper.
By now, it has been established in a solid way that only quasi-conformal technicolor
 theories can pass the electroweak precision tests and simultaneously avoid the problems of old technicolor theories (like large FCNC). To our knowledge the type of technicolor models we examine are the only viable technicolor models that not only are not excluded, but they have desired features (light Higgs, stable DM etc.).
The model we have focused, MWT is the simplest of all: 2 techniquarks in the adjoint of SU(2).
Our analysis is valid for general TC DM that are not based on the stability of technibaryon number
or on excess of particle over antiparticle.

The paper is organized as follows: In section 2, we calculate the relevant annihilation cross sections
for \N2 in the early Universe and the Sun. In section 3, we calculate the capture rate of relic \N2
 by the Sun and the Earth. In section 4, we provide the muon flux from the captured dark matter particles.
 We present our conclusions and a discussion of the uncertainties that might enter in our results in section 5.

\section{Annihilation of \N2 in the early Universe and in the Sun}

A pair of \N2 annihilates mainly into pairs of fermions $f\bar{f}$ and into pairs of W-bosons $W^+W^-$
(with longitudinal polarization), provided that the energy is sufficient to open the corresponding channel.
In our calculation we adopt the following formulas for the annihilation cross sections
multiplied by the relative velocity and averaged over the thermal velocity distribution at temperature $T$
\beq
\sv_{ff} = \frac{2G_F^2m^2\beta_f}{\pi}P_Z\left\{\frac{C_A^2}{2}\frac{m_f^2}{m^2}+ \left[(C_V^2+C_A^2)+
\left(\frac{C_V^2}{2}-\frac{17C_A^2}{8}\right)\frac{m_f^2}{m^2}\right]\frac{T}{m}\right\}\sin^4\theta,
\label{ff}
\eeq
\beq
\sv_{WW} = \frac{2G_F^2m^2(2m^2-m_W^2)^2\beta_W^3}{\pi m_Z^4}P_Z\frac{T}{m}\sin^4\theta,
\label{WW}
\eeq
 which are deduced from \cite{CK, fin, Kolb, Gaisser}.
  Here
$$P_Z=\frac{m_Z^4}{(4m^2-m_Z^2)^2+\Gamma_Z^2m_Z^2},$$
$$\beta_{f,W}=\sqrt{1-m_{f,W}^2/m^2},$$
where $m_f$, $m_Z$, $m_W$, and $m$ are the masses of the final fermion $f$, $Z$-, $W$-bosons, and $N_2$ respectively, $G_F$ is the Fermi constant,
$C_V=T_{3L}-2Q\sin^2\theta_W$, and $C_A=T_{3L}$ are Standard Model parameters of $f$. $T_{3L}$ and $Q$ are the weak isospin and the electric charge of the corresponding particle.
Eqs.~(\ref{ff},~\ref{WW}) represent the cross sections in a non-relativistic approximation in the form of
\beq
\sv=\sigma_0+\sigma_1\frac{T}{m}.
\label{sv_series}
\eeq
In the Standard Big Bang scenario, the modern relic density of \N2 is given by \cite{CK}
\beq
\Omega_{N_2}h^2=\frac{1.76\cdot 10^{-10}\sqrt{g_*}}{g_{*s}\sigma_1[\GeV^{-2}]}\fr{m}{T_*}{2},
\label{Omega}
\eeq
where the freeze out temperature is given by
\beq
\frac{m}{T_*}=L-\frac{3}{2}\ln [L ],
\qquad L=\ln \left[\frac{m_{Pl}m\sigma_1}{6.5\sqrt{g_*}}\right].
\label{xf}
\eeq
 $m_{Pl}$ is the Planck mass, $g_*$ and $g_{*s}$ are effective spin degrees of freedom contributing into the energy and entropy densities of the plasma at $T=T_*$ respectively. For the values of interest of $T_*$, $g_*=g_{*s}=80\div 100$.

By inspection of Eqs.~(\ref{ff},~\ref{WW}), we see that at the freeze-out, the cross section is dominated by $\sv=\sigma_1\cdot T/m$
for all the annihilation channels ($N_2 N_2\to WW$, $N_2 N_2\to f\bar{f}$, with $f$ being $\nu_{e,\mu,\tau},e,\mu,\tau$,
 $u,d,s,c,b$). For the annihilation of \N2 into pairs of top quark-antiquark, the first term
  of Eq.~(\ref{sv_series}) can be important (for the mass interval $m_t<m\lsim 500$GeV),
   however for $m>m_t$, the $WW$-channel prevails strongly in the period of freeze-out.

\begin{figure}
\begin{center}
\includegraphics[scale=0.6]{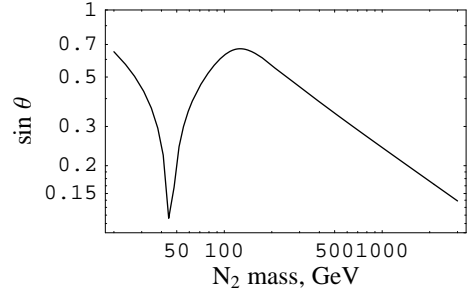}
\caption{Dependence of the mixing parameter $\sin \theta$ on the mass of \N2.}
\label{sin}
\end{center}
\end{figure}

\begin{figure}
\begin{tabular}{cc}
\mbox{\includegraphics[scale=0.5]{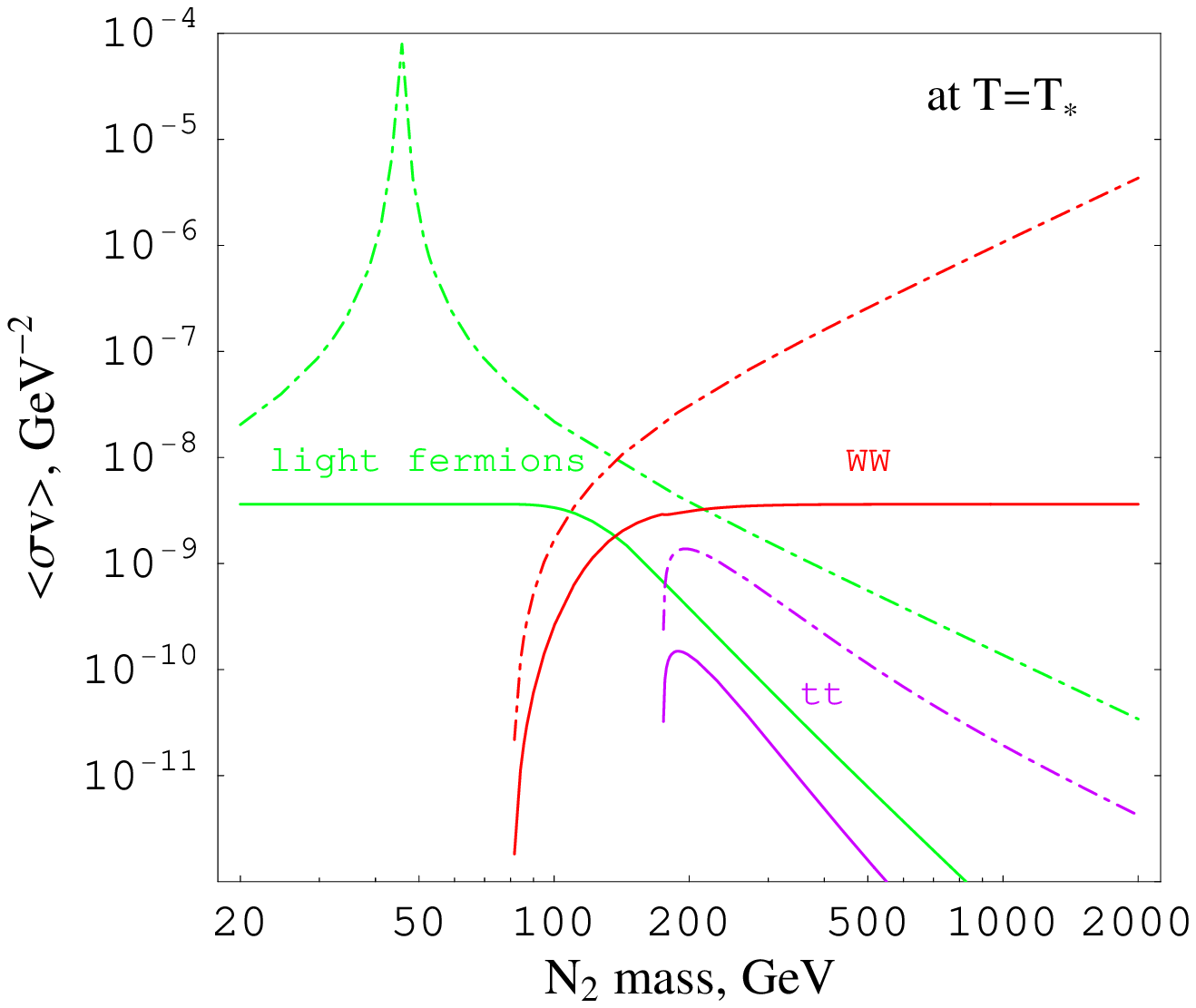}}&
\mbox{\includegraphics[scale=0.5]{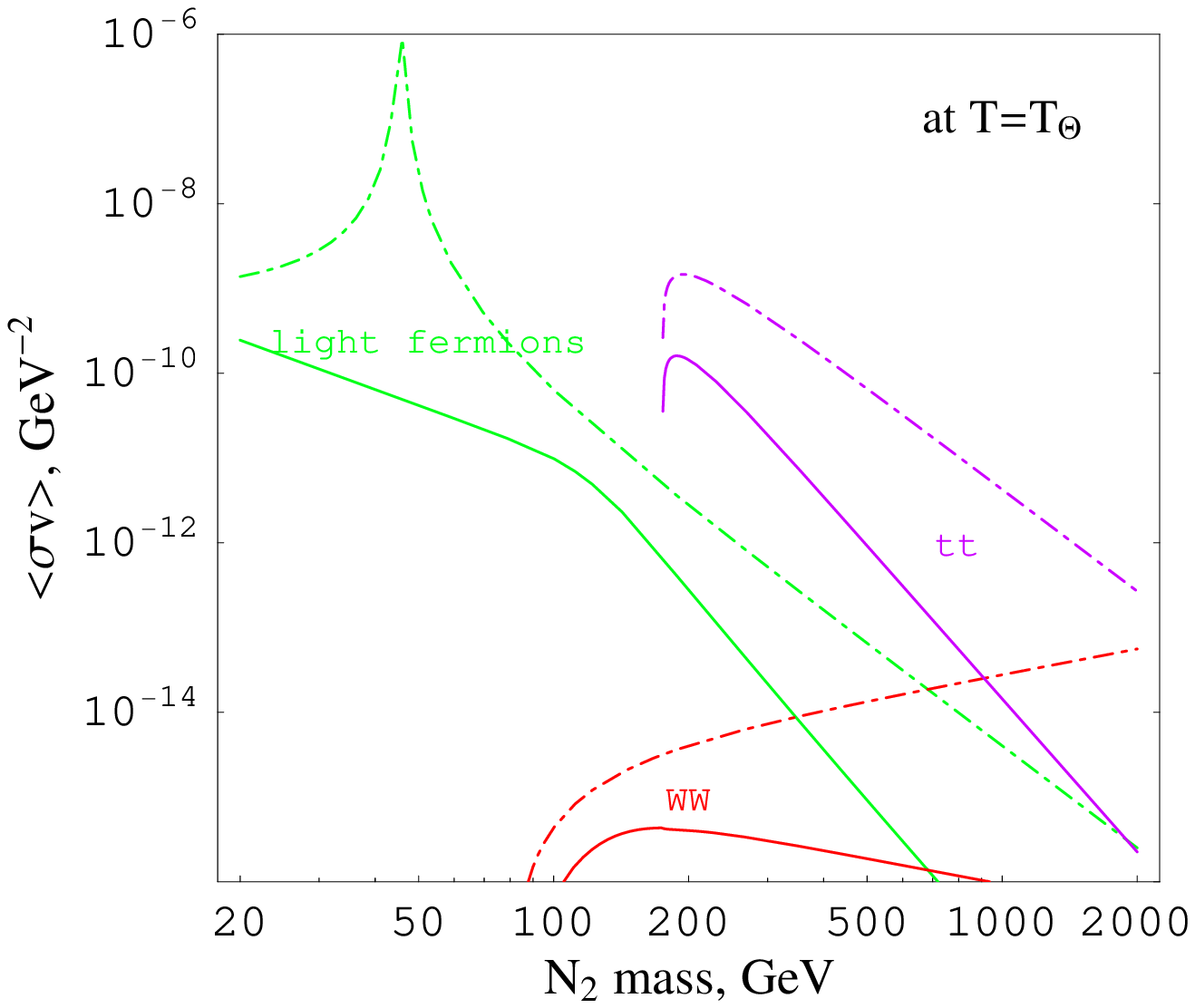}}
\end{tabular}
\caption{The cross sections of $N_2N_2\to$ light fermions, $t\bar{t}, WW$ are shown for the models \cite{CK} (solid lines) and \cite{fin} (dot-dashed lines) at freeze-out (left) and in the Sun (right). ``Light fermions'' include all fermions except the top-quark; channels with off-mass shell $W$ or $f$ were not taken into account.}
\label{cross}
\end{figure}

The condition that \N2 saturates all the Cold Dark Matter (CDM)
\beq
\Omega_{N_2}h^2=\Omega_{\rm CDM}h^2=0.112
\label{omega_CDM}
\eeq
fixes the free parameter of the model $\sin\theta$ (or $\sigma_1$ in Eqs.~(\ref{Omega},~\ref{xf})). It is given on Fig.~(\ref{sin}) as function of the mass of $N_2$. In the model of pure Majorana neutrino \N2~(i.e.\ $\nu'$) \cite{fin}, where $\sin\theta=1$, such a condition is provided by an extension of the Big Bang scenario due to quintessence~\cite{kination}. Due to this, there is a difference between the annihilation cross sections predicted in these two variants of MWT DM particles as shown on Fig.~(\ref{cross}, left). In the model~\cite{CK}, the total cross section at the freeze-out is virtually independent of the mass due to the fact that it is always adjusted via Eq.~(\ref{omega_CDM}) in order to give the proper relic density.
However, at the temperatures of the solar core, $T_{\odot}\approx 1.3\keV\ll T_*$, 
  the first term in Eq.~(\ref{sv_series}) (or Eq.~(\ref{ff})) dominates for many types of final fermions. Therefore annihilation rates inside the Sun are different from those in the early Universe (see Fig.~(\ref{cross}, right)). Channels with light final fermions in the Sun become suppressed with respect to those in the early Universe, while the $t\bar{t}$-channel becomes of special importance (see section \ref{MuonFlux}).

\section{Capture of relic \N2 by the Sun and the Earth}

Relic \N2 with density in the vicinity of the solar system assumed to be
$\rho_{loc}= 0.3 \Gcm$,
may scatter off nuclei inside the Sun and the Earth and can be trapped by their corresponding gravitational potential wells. The interaction of \N2 with nuclei $A$ is spin dependent and the respective cross section can be represented as \cite{CK}
\beq
\sigma_{N_2A} = \frac{2G_F^2\mu^2}{\pi}I_s\sin^4\theta,
\label{sigma}
\eeq
where $\mu$ is the the reduced mass of \N2 and $A$, and
\beq
I_s=C^2\cdot \lambda^2J(J+1).\\
\label{Is}
\eeq
The coefficient $C$ takes into account quark contributions to the spin of the nucleon and for weak interaction is \cite{Lewin,Ellis}
\beq
C=\sum_{q=u,d,s} T_{3q}\Delta q \approx\left\{
\begin{array}{ll}
\frac{1}{2}0.78-\frac{1}{2}(-0.48)-\frac{1}{2}(-0.15)=0.705
& \textrm{for }p\\
\frac{1}{2}(-0.48)-\frac{1}{2}0.78-\frac{1}{2}(-0.15)=-0.555
& \textrm{for }n.
\end{array} \right.
\label{C}
\eeq
The other coefficient in Eq.~(\ref{Is}) relates the nucleon contribution (with spin $s$ and orbital momentum $l$) to the spin $J$ of the nucleus, and within the single unpaired nucleon model it is
\beq
\lambda^2J(J+1)=\frac{[J(J+1)+s(s+1)-l(l+1)]^2}{4J(J+1)}.
\label{lambdaJ}
\eeq
We assume that the hydrogen is the only one to contribute to the capture of relic \N2 in the Sun. In this case from Eqs.~(\ref{Is}--\ref{lambdaJ}), we have $I_s=0.705\cdot\frac{3}{4}\approx 0.37.$ This estimate agrees with that for Dirac neutrino taking into account only the axial current (spin-dependent) contribution, which gives \cite{4N} $I_s\approx 1.3^2\cdot 3/16\approx 0.3$.
The case of the Earth will be commented separately.

\begin{figure}
\begin{tabular}{cc}
\mbox{\includegraphics[scale=0.52]{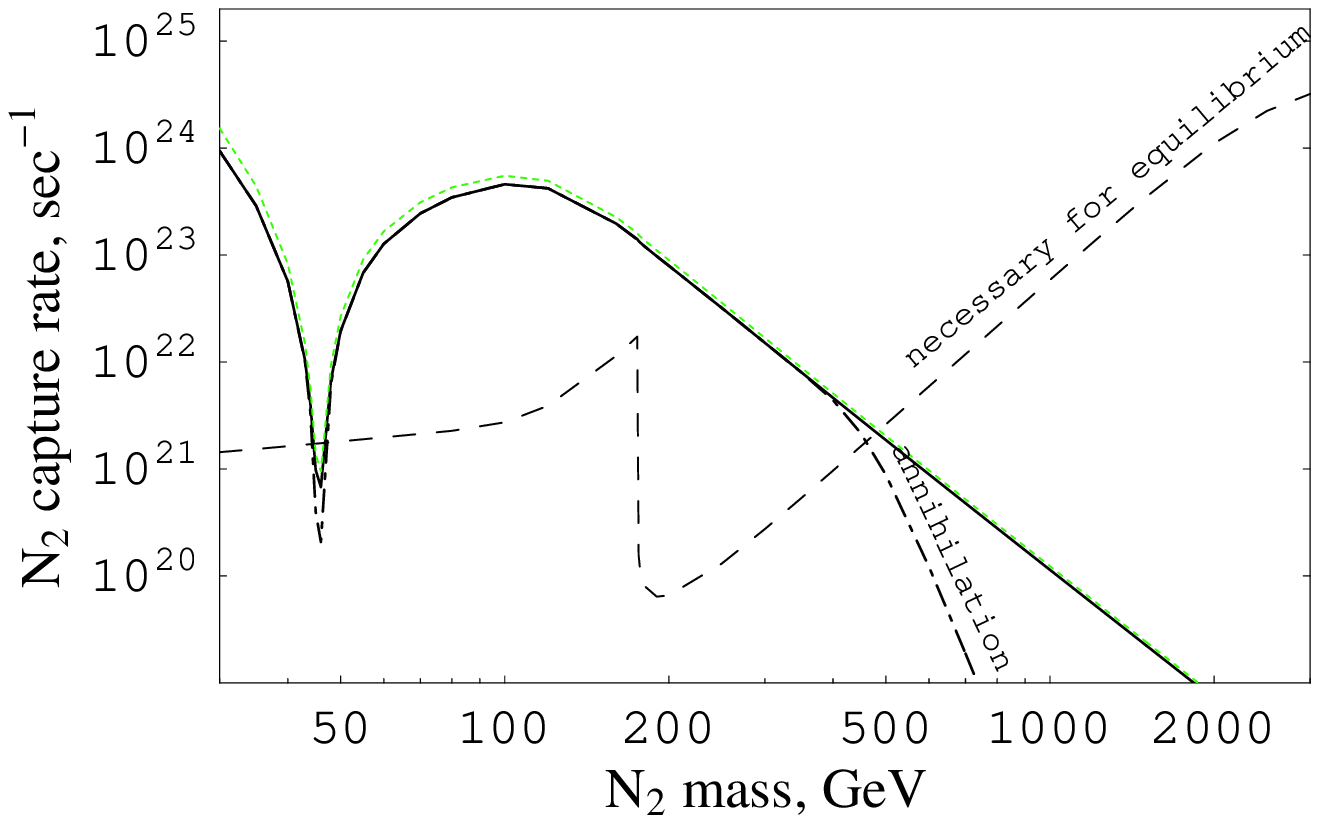}}&
\mbox{\includegraphics[scale=0.52]{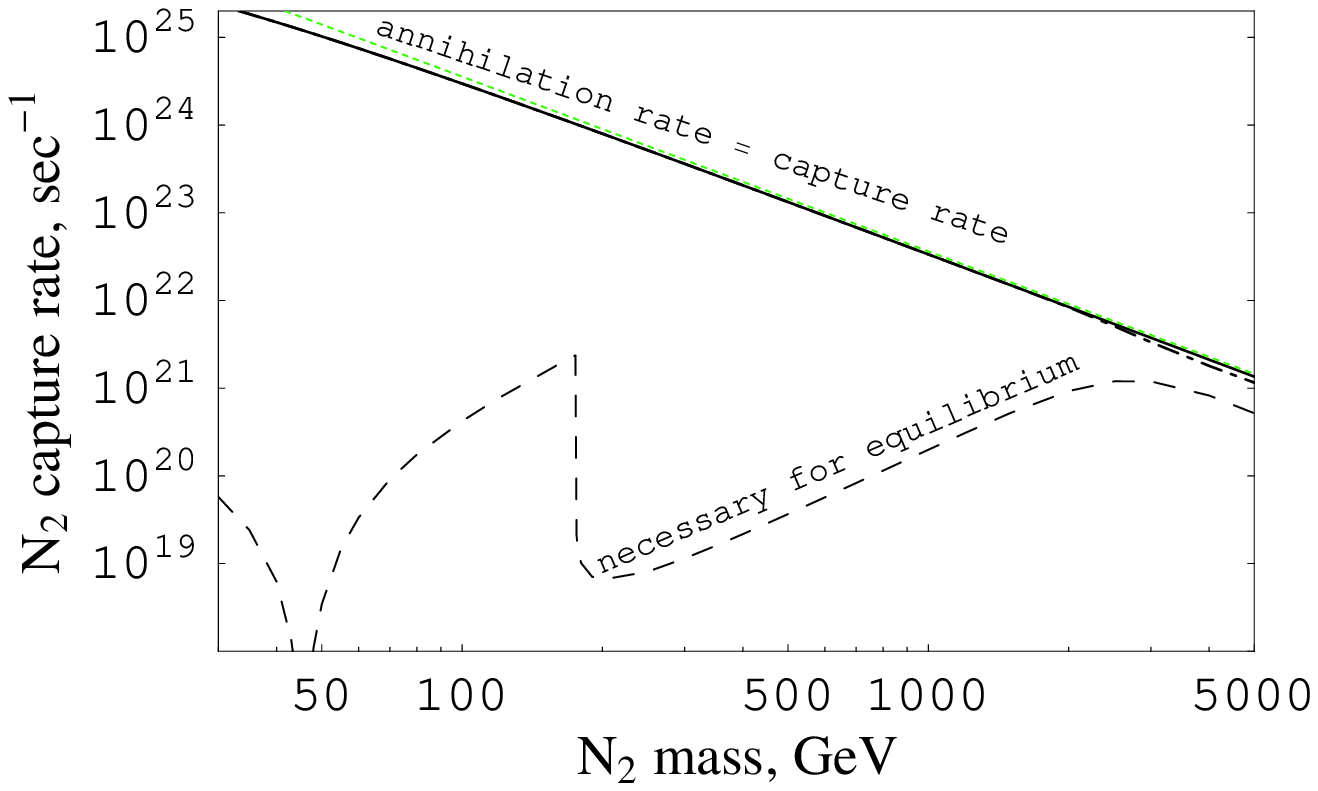}}
\end{tabular}
\caption{The capture and annihilation rates for \N2 in the case of the Sun for the models \cite{CK} (left) and \cite{fin} (right). Dot light (green) lines show the capture rates obtained in the approximation \cite{Gould, Jungman}.}
\label{Ncapann}
\end{figure}

It is not difficult to estimate the capture rate of \N2 by the Sun or the Earth. The common expression for that is
\beq
\dot{N}_{capt}=\sum_A \int n_{N_2}\langle \sigma'_{N_2A}v\rangle n_A dV,
\label{Ncap}
\eeq
where $n_{N_2}$ and $n_A$ are the number densities of \N2 and $A$ in a given volume element $dV$,  $\sigma'_{N_2A}$ is the cross section for a \N2-$A$ collision times the probability that \N2 loses enough energy to be trapped gravitationally by the Sun. Introducing a nuclear form factor $F_A$, one writes
\beq
\sigma'_{N_2A}=\sigma_{N_2A}\int^{\DT_{\max}}_{T_{\infty}}F^2_A(\DT)\frac{d\DT}{\DT_{\max}}=
\sigma_{N_2A}\,\overline{F^2_A}\,\frac{v^2_{esc}(r)-\delta v^2_{\infty}}{v^2},
\label{sigmaprime}
\eeq
where $\DT$ is the transferred energy in a \N2-$A$ collision, $\DT_{\max}=2\mu^2v^2/m_A$, $\delta=(m-m_A)^2/(4m m_A)$ with $m_A$ being the nucleus mass, $v_{esc}$ is the escape velocity at distance $r$ from the center of gravity, $v=\sqrt{v_{\infty}^2+v_{esc}^2}$ and $v_{\infty}$ are the \N2 velocities at distances $r$ and $r\to \infty$ respectively. $\overline{F^2_A}$ is the $F_A^2$ averaged over the interval $\Delta T \in [T_{\infty}; \Delta T_{\max}]$, where $T_{\infty}\equiv mv_{\infty}^2/2$.
The number density $n_{N_2}$ of \N2 at distance $r$, can be related with the one outside the potential well, $n_{N_2}(r\to \infty)\equiv n_{N_2\,\infty}$, as \cite{4N,nv}
$n_{N_2}=n_{N_2\,\infty}\cdot v/v_{\infty}.$
We average Eq.~(\ref{Ncap}) over the velocity distribution, by the substitution
$$n_{N_2\,\infty}\to n_{N_2\,\infty}\cdot f_{\infty}(v_{\infty})dv_{\infty},$$
where $n_{N_2\,\infty}=\rho_{loc}/m$. We use a velocity distribution of the form
\beq
f_{\infty}(v_{\infty})=\frac{v_{\infty}}{\sqrt{\pi}v_0v_{\odot}}
\left(\exp\left[-\frac{(v_{\infty}-v_{\odot})^2}{v_0^2}\right]-\exp\left[-\frac{(v_{\infty}+v_{\odot})^2}
{v_0^2}\right]\right),
\label{fv}
\eeq
where $v_{\odot}=v_0=220$ km/s. 
The captured \N2 accumulate in the solar core and annihilate. Their number density is governed by the equation
\beq
\dot{N}=\dot{N}_{capt}-\dot{N}_{ann}.
\label{dotN}
\eeq
Here, $\dot{N}_{ann}$ is the number of \N2 disappearing due to annihilation per second,
$$\dot{N}_{ann}=\int n_{N_2}^2\sv dV.$$
It is two times larger than the rate of annihilation acts. The effect of evaporation of the captured
and thermalized \N2 is neglected, something that is valid for all $m\gsim 3$ GeV \cite{Gaisser}.
Thermalization of the captured \N2 happens, due to succession of collisions with nuclei, well before \N2 has time to annihilate (the ratio of respective characteristic times, within \N2 mass range of interest, is $\sim 10^{-5\div -10}$ in case of Sun).
Resolving Eq.~(\ref{dotN}) for $\dot{N}_{ann}$, one finds
$$\dot{N}_{ann}=\dot{N}_{capt}\tanh^2\left[ \sqrt{\frac{\dot{N}_{capt}}{\dot{N}_{eq}}}\right].$$
Here $$\dot{N}_{eq}=\frac{V_{therm}}{\sv t_{age}^2}$$
defines a critical capture rate above which equilibrium between capture and annihilation is established during the solar lifetime $t_{age}$. For $\dot{N}_{capt}\ll \dot{N}_{eq}$, $\dot{N}_{ann}$ is suppressed with respect to $\dot{N}_{capt}$ as $\dot{N}_{capt}/\dot{N}_{eq}$. The value
\begin{eqnarray}
\lefteqn{V_{therm}=\left(\frac{4\pi\bar{\rho}}{\rho_{core}}\frac{T_{core}}{T_{esc}}\right)^{3/2}\! R^3
\approx \fr{\rm TeV}{m}{3/2}\times} \nonumber\\
& & \times \left\{
\begin{array}{ll}
2.0\cdot 10^{26}\cm^3\fr{T_{core}}{15\cdot 10^6 \K}{3/2}\fr{150\gcm}{\rho_{core}}{3/2}
& \textrm{for the Sun,}\\
1.0\cdot 10^{23}\cm^3\fr{T_{core}}{7000\K}{3/2}\fr{11\gcm}{\rho_{core}}{3/2}
& \textrm{for the Earth,}
\end{array} \right.
\label{V}
\end{eqnarray}
characterizes the effective volume that the captured \N2 occupy, after being thermalized, having a Maxwell-Boltzmann velocity distribution. Here, $R$ is the radius of the Sun or the Earth, $\bar{\rho}$, $\rho_{core}$ and $T_{core}$ are their mean and core densities and core temperatures respectively (for the Sun $T_{core}\equiv T_{\odot}$), $T_{esc}\equiv mv_{esc}^2(r=R)/2$. For the derivation of Eq.~(\ref{V}) we assumed that the density of matter and the temperature within $V_{therm}$ are constant and equal to their core values. In this case, the potential energy with respect to the center takes the form $U(r)=T_{esc}(\rho_{core}/\bar{\rho})(r/R)^2/2$, and an integration of the thermalized \N2 number density, $n_{N_2}(r)=n_{N_2}(0)\exp(-U(r)/T_{core})$, can be done analytically.
 Note, that the quantity given in Eq.~(\ref{V}) for the Sun agrees with the one in \cite{Silk}.
For the integration in Eq.~(\ref{Ncap}), we assume a matter density distribution in $r$ as in \cite{4N}. The effect of the finite size of hydrogen is insignificant in this case. In fact, $qa<0.1$ (typically $\sim 0.02$), where $q=\sqrt{2m_A\Delta T}$ is the transferred 3-momentum and $a$ is the nucleus size, so $F_A(qa)\approx 1$. The capture and annihilation rates obtained for the case of the Sun are shown in Fig.~(\ref{Ncapann}).
For comparison, the same results obtained within the approximation of \cite{Gould, Jungman}
$$\dot{N}_{capt}\approx 4.5\cdot 10^{18}\s\cdot \frac{\rho_{loc}}{0.4\Gcm}\fr{270\kms}{\bar{v}}{3}
\frac{\sigma_{H,SD}}{10^{-6}\pb}\fr{1000\GeV}{m}{2}$$
are shown in Fig.~(\ref{Ncapann}) too. As seen, the agreement is very good.

\begin{figure}
\begin{tabular}{cc}
\mbox{\includegraphics[scale=0.5]{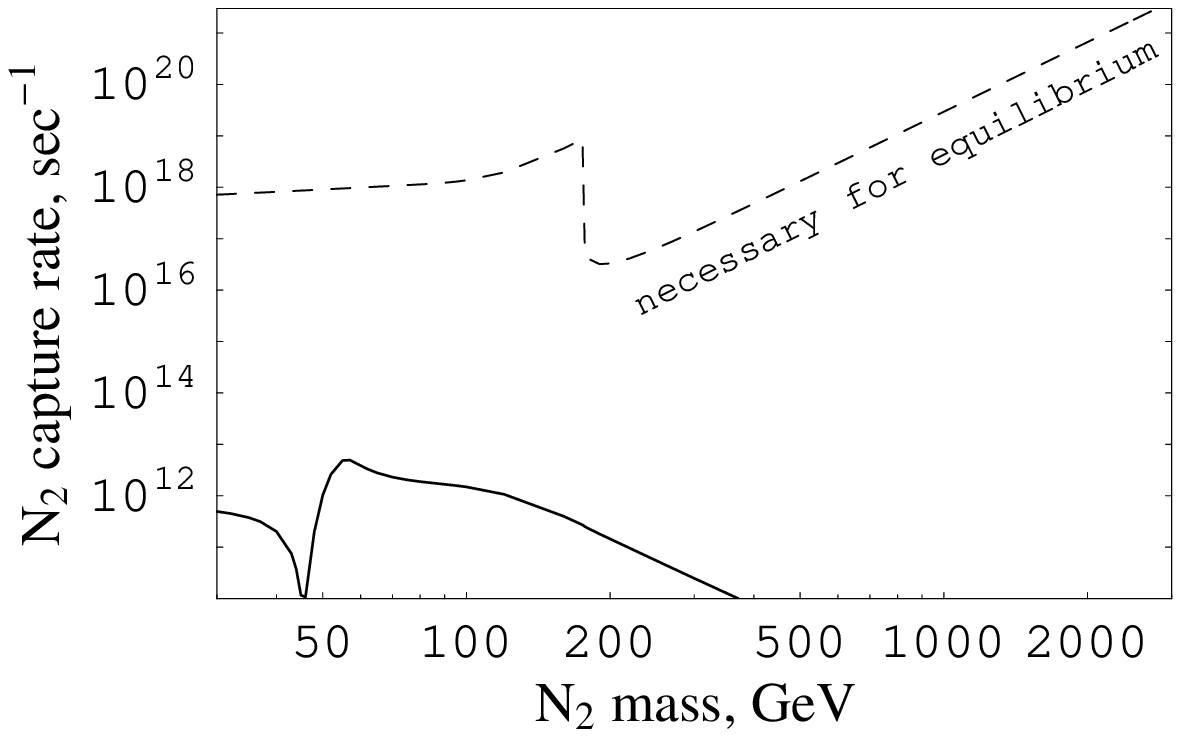}}&
\mbox{\includegraphics[scale=0.5]{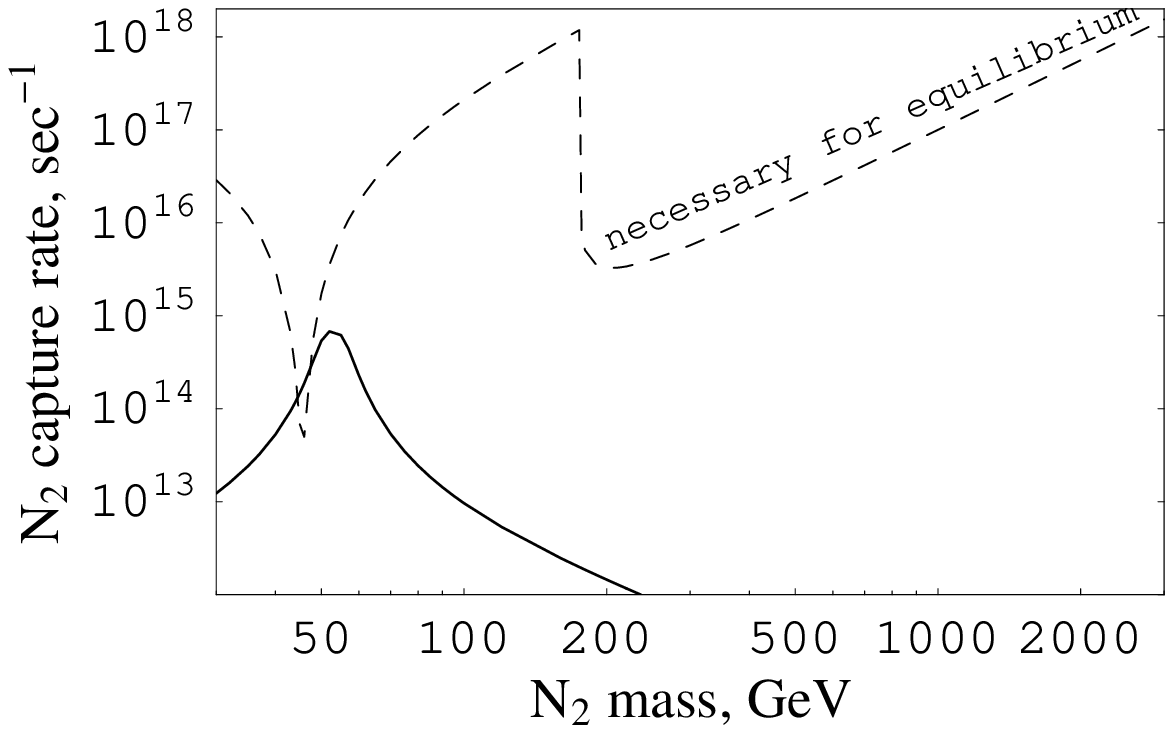}}
\end{tabular}
\caption{A rough estimate of the capture rates of relic \N2 by the Earth for the models \cite{CK} (left) and \cite{fin} (right).}
\label{NcapE}
\end{figure}

As to the capture by the Earth, the essential difference is that the potential well in this case is very ``shallow'', and incident DM particles have a chance to be captured only at special kinematic conditions. It may happen in scattering off nuclei with nonzero spin, if the DM particle has a mass close to the mass of the nucleus or/and is initially very slow. As a rough estimate, we take one of the nucleus with nonzero spin being quite abundant in the Earth: the isotope $^{57}Fe$ present in natural $Fe$ with fraction $\approx 2$\%, while all iron is assumed to make up $\approx 30\div 40$\% of the Earth's mass. We assume that it is concentrated in the core of the Earth. In the core $v_{esc}(r\approx 0)\approx 14.2$ km/s. The cross section of the $^{57}Fe$-\N2 interaction is defined in Eq.~(\ref{sigma}). For $I_s$, one can give a maximal estimate in the single unpaired nucleon approximation of $I_s\approx 0.23$ (from Eqs.~(\ref{Is}--\ref{lambdaJ})). The form factor of iron can be roughly estimated in the thin sphere approximation \cite{Lewin}
$$F_A(qa)=\frac{\sin(qa)}{qa}.$$
Since the integration interval in Eq.~(\ref{sigmaprime}) is small (in the case of the Earth) with respect to the characteristic scale of the $F_A$ variation, we take $\overline{F^2_A}=F^2_A(T_{\infty})$.
For a simple (maximal) estimate, we use Eq.~(\ref{fv}) as the \N2 distribution in the vicinity of the Earth, where the depletion in the $v$-space for $v<42$ km/s caused by solar attraction is ignored (moreover, we also neglect the effect of possible accumulation of dark matter particles in the solar system, described in \cite{Damour}). The capture rates of \N2 by the Earth's potential due to $^{57}Fe$-\N2 collisions as obtained for the two MWT considered models are represented in Fig.~(\ref{NcapE}).
As seen in Fig.~(\ref{Ncapann},~\ref{NcapE}), in the model \cite{CK}, $\dot{N}_{capt}$ is lower and $\dot{N}_{eq}$ is higher (compared to other considered model), which is a consequence of the $\sin^4{\theta}$ suppression of both \N2\N2 and $A$-\N2 interactions (Eqs.~(\ref{ff},~\ref{WW},~\ref{sigma})). In the case of the Earth, the annihilation channel $N_2N_2\to WW$ is suppressed with respect to the case of the Sun by $7000\K/15\cdot 10^6\K\sim 5\cdot 10^{-4}$, and therefore $\dot{N}_{eq}$ in Fig.~(\ref{NcapE}) does not fall for $m\gsim 2$ TeV. Similarly, neutrino yields from \N2 annihilation in the Earth and the Sun should not differ for $m\lsim 2$ TeV, except for a difference caused by absorbtion effects in the solar matter, being meaningful only for high \N2 masses.

 By comparison of Figs.~(\ref{Ncapann},~\ref{NcapE}), we see that for the maxima of $\dot{N}_{capt}$ in the case of the Earth, which occur for a mass of $N_2$ close to the mass of $^{57}Fe$, the ratio of $\dot{N}_{capt}$ for the Earth over the Sun is $\sim 10^{-10}$ in both considered models~\cite{CK,fin}.
 The neutrino flux, induced by annihilation of \N2 in the Earth, will differ from that from the Sun as the aforementioned ratio multiplied by the squared ratio of distances to the Sun and the Earth centers $5\cdot 10^8$
and a factor $\dot{N}_{capt}/\dot{N}_{eq}\sim 10^{-4}$ for the model \cite{CK}.
 So, even for the model \cite{fin}, where the $\dot{N}_{capt}/\dot{N}_{eq}$ suppression is much weaker, the \N2 annihilation induced neutrino flux from the Earth at its maximum is a few tens times less than that from the Sun. A greater neutrino flux from the Earth relatively to that from the Sun can be hardly expected from collisions of relic \N2 with other nuclei present in the Earth, which the abundance is more uncertain. Since the sensitivity of the Super-Kamiokande experiment to the neutrino induced muon flux from both the Solar and Earth cores is of the same order of magnitude, we shall neglect the \N2 annihilation effects in the Earth in the following consideration.

\section{Muon flux from the captured \N2}
\label{MuonFlux}

Annihilation of \N2 produces $e$-, $\mu$-, and $\tau$-neutrinos with a flux
$$\Phi_{\nu}=\dot{N}_{ann}\frac{N_{\nu}}{2\cdot4\pi r^2},$$
where $N_{\nu}$ is the multiplicity of neutrinos produced per \N2\N2-annihilation, and $r$ is the distance to the center of the Sun or the Earth. Neutrinos from annihilation passing through the solar matter can reach the Earth, traverse it and induce at its surface the muon flux
\beq
\Phi_{\mu}=\Phi_{\nm}\frac{x_{\mu}}{x_{\nm}}
=\frac{\dot{N}_{ann}}{2\cdot4\pi r^2}\int dN_{\nm}(E_{\nm})\frac{\aver{x_{\mu}}}{x_{\nm}} ,
\label{Fmu0}
\eeq
where $x_{\mu}$ and $x_{\nm}$ are the mean free paths (measured as matter columns in g/cm$^2$) of $\mu$ and $\nm$ with respect to the processes of energy loss (mainly ionization) and $\nm+A\to \mu+X$ respectively.
The last equality in Eq.~(\ref{Fmu0}) is a generalization for the case of energy dependence.

For muon energy losses we use the approximation \cite{Review}
\beq
-\frac{dE_{\mu}}{dx_{\mu}}=a\approx 2.3\frac{\rm MeV}{\rm g/cm^2},
\label{dEdx}
\eeq
taking into account ionization effects and ignoring pair production and bremsstrahlung effects.  This approximation
 is valid for $E_{\mu}\lsim 800\GeV$ and seems reasonable within the energy interval of question
  as will be seen from the final result.
The muon mean free path is
$$x_{\mu}(E_{\mu})=E_{\mu}/a.$$
For the muon mean energy $\aver{E_{\mu}}$ in the reaction $\nm+A\to \mu+X$, defining $\aver{x_{\mu}}$, we use following \cite{Jungman,Ritz}, the relationship
\beq
\aver{E_{\mu}}=b\cdot E_{\nm},\qquad b= \left\{
\begin{array}{ll}
0.5 & \textrm{for }\nm,\\
0.7 & \textrm{for }\anm.
\end{array} \right.
\label{b}
\eeq
The neutrino mean free path is governed by the charged current interaction with nucleons. The corresponding cross section calculated in \cite{Gandhi}, can be approximated within accuracy $\sim 10$\% as
\beq
\sigma_{CC}\approx \left\{
\begin{array}{ll}
0.72\cdot 10^{-36}\frac{E_{\nm}}{E_0}\cm^2 & \textrm{for }\nm,\\
0.37\cdot 10^{-36}\frac{E_{\nm}}{E_0}\cm^2 & \textrm{for }\anm,
\end{array} \right.
\label{sigmaCC}
\eeq
where $E_0=100$ GeV. For the respective mean free path one has
\beq
x_{\nm}=\frac{1}{N_A\sigma_{CC}}\approx x_0\frac{E_0}{E_{\nm}}, \quad
x_0\approx \left\{
\begin{array}{ll}
2.3\cdot 10^{12}\gc & \textrm{for }\nm,\\
4.5\cdot 10^{12}\gc & \textrm{for }\anm,
\end{array} \right.
\eeq
where $N_A=6\cdot 10^{23}$ 1/g is the number of nucleons per gram of matter.
Separating annihilation neutrinos from different channels $ch$ in Eq.~(\ref{Fmu0}), we have
\beq
\Phi_{\mu}=\frac{\dot{N}_{ann}}{8\pi r^2}\frac{b}{ax_0E_0}
\sum_{ch} Br_{ch}N_{\nm(ch)}\aver{E^2_{\nm(ch)}},
\label{Fmu1}
\eeq
where $Br_{ch}$, $N_{\nm(ch)}$ and $\aver{E^2_{\nm(ch)}}$ are respectively the branching ratio, the neutrino yield and the mean neutrino energy for a given channel. Eq.~(\ref{Fmu1}) agrees numerically within $20\div 40$\% with the respective formula of \cite{Jungman}.
\begin{figure}
\begin{tabular}{cc}
\mbox{\includegraphics[scale=0.5]{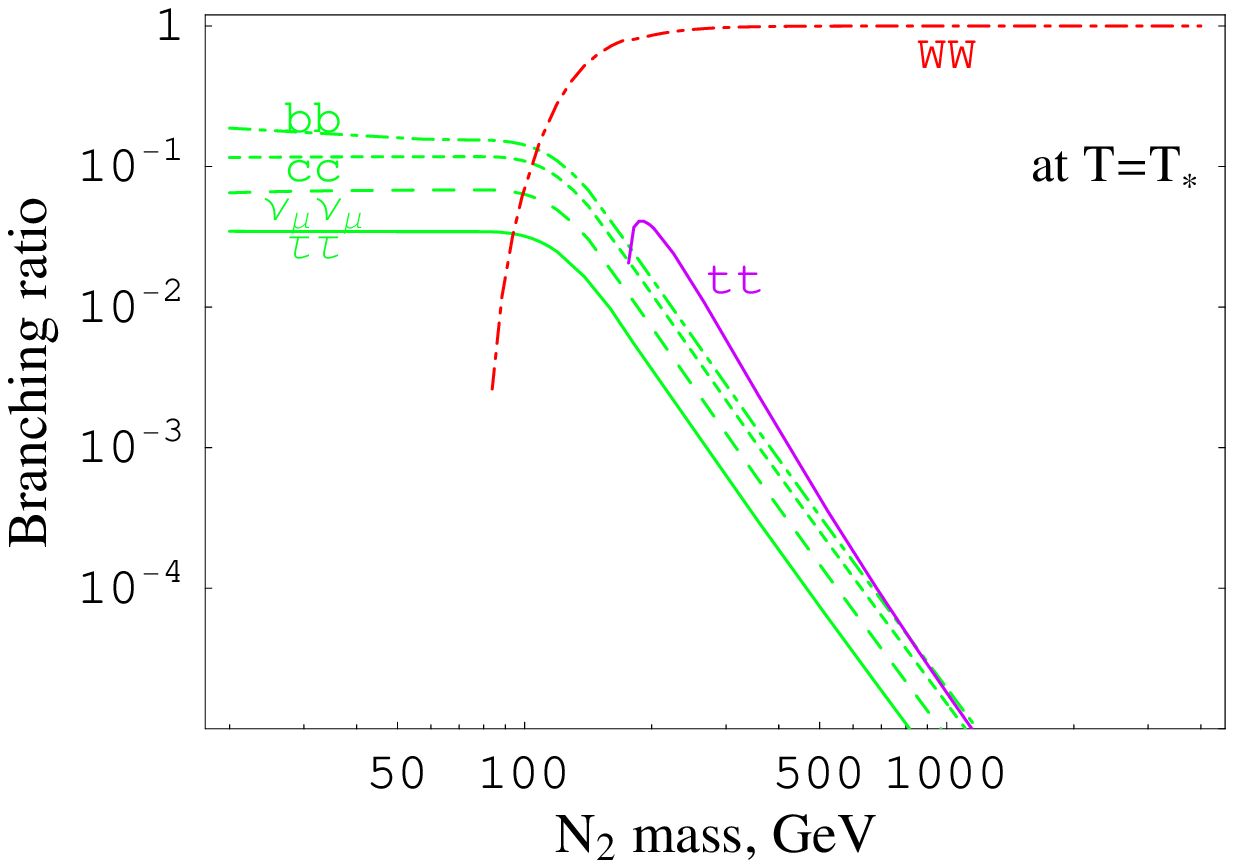}}&
\mbox{\includegraphics[scale=0.5]{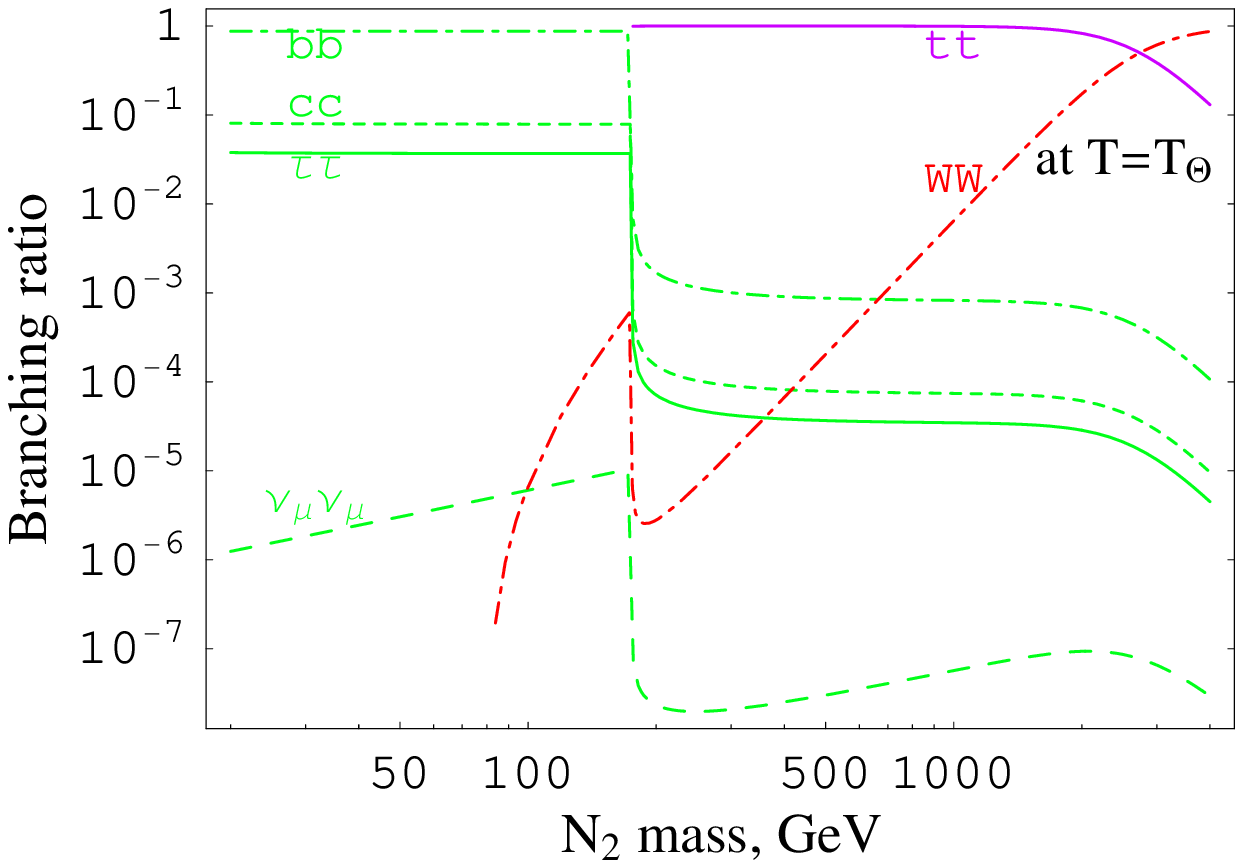}}
\end{tabular}
\caption{The branching ratios of the most interesting \N2 annihilation channels for the cases of freeze-out (left) and in the solar core (right). The models \cite{CK} and \cite{fin} do not differ in this plot.}
\label{Br}
\end{figure}

Fig.~(\ref{Br}) illustrates the difference in the most important \N2 annihilation channels for
the cases of freeze-out (left) and solar core (right).
Muon neutrinos produced in the channels $N_2N_2\to \bar{u}u,\bar{d}d,\bar{s}s$ as well as in decays of any born muons are not of interest because the primary particles have enough time to slow down before they decay, preventing neutrinos from producing a signal above the experimental threshold. Also, in the dense solar core, $c$- and $b$-quarks with initial energy $\gsim 100$ GeV lose partially their energy. However, we shall neglect this effect. It is not expected to cause an essential error, because as we will show below, the muon signal is predicted to be virtually unobservable when highly energetic $c$- and $b$-quarks are born as a result of \N2 annihilation. In fact, for large \N2 mass, the $\bar{t}t$- and $WW$-channels (see Fig.~(\ref{Br})) dominate. Decaying, they give $c$- and $b$-quarks with degraded energy. So, a very large mass of \N2 is needed to noticeably produce highly energetic $c$- and $b$-quarks. However, for such masses, the muon signal is predicted to be small because of suppression of the $\dot{N}_{ann}$ itself (in the model \cite{CK}) or/and because of effects of $\nm$ absorbtion in the solar matter. Generally,
as seen in Eq.~(\ref{Fmu1}), in the absence of $\nm$ absorbtion in the matter, a decrease of neutrino energy reduces quadratically the intensity of the muon signal. For this reason, channels giving rise to $\nm$ as a result of a long cascade chain, can be considered to be negligible with respect to similar channels with shorter cascade chains of $\nm$ production. Moreover, as a thumb rule, longer chains have an additional suppression due to the branching ratio. For example, the channels $N_2N_2\to WW\to \tau\nu, bc, cs \to \nm X$ have a suppression with respect to $N_2N_2\to WW\to \mu\nm$  $\sim 10^2$ times.

\begin{figure}
\begin{tabular}{cc}
\mbox{\includegraphics[scale=0.5]{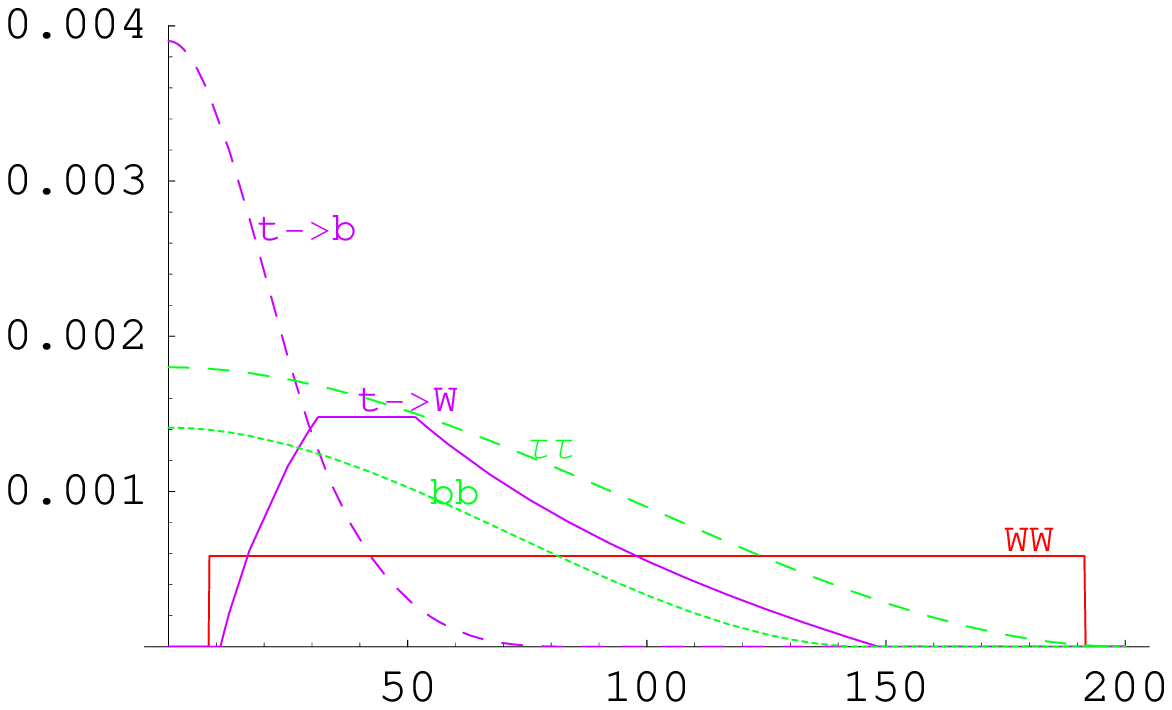}}&
\mbox{\includegraphics[scale=0.5]{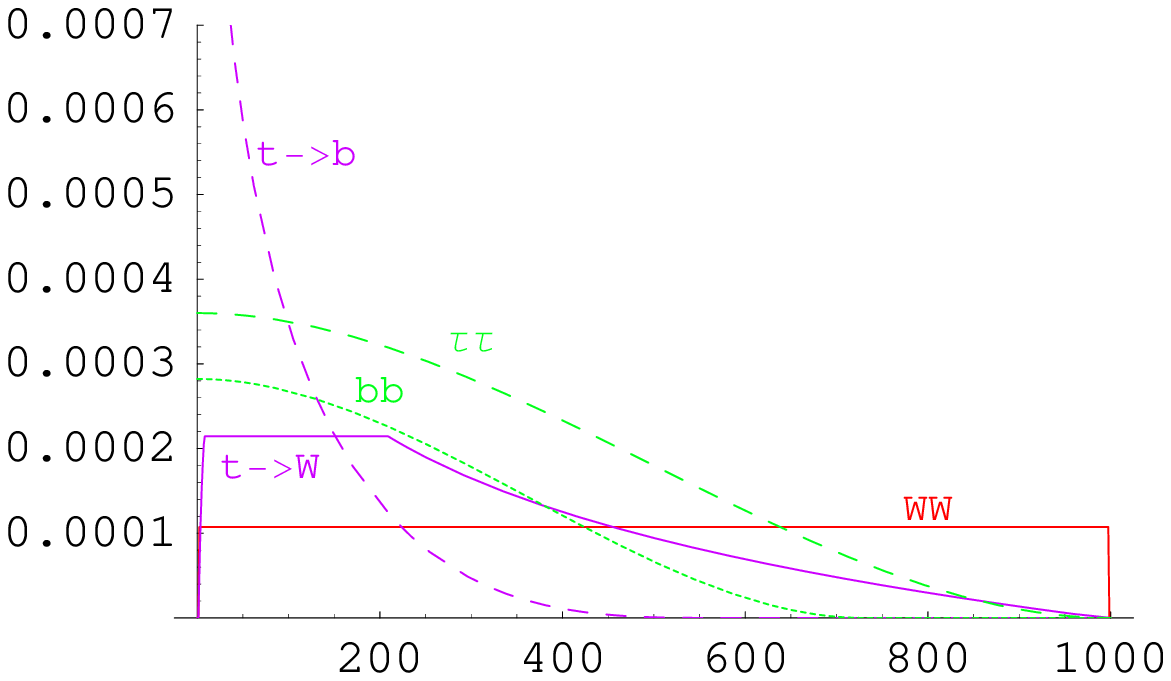}}
\end{tabular}
\caption{The energy spectra of $\nm$ produced in different channels for $m=200$ GeV (left) and $m=1000$ GeV (right).}
\label{spectra}
\end{figure}

We shall take into account the following channels
\beq
N_2N_2\to \nm\anm,
\label{Ntonu}
\eeq
\beq
N_2N_2\to \tau^+\tau^-\to \mu \anm \nu_{\tau},
\label{Ntotau}
\eeq
\beq
N_2N_2\to c\bar{c}\to \mu\anm X,
\label{Ntoc}
\eeq
\beq
N_2N_2\to b\bar{b}\to \mu\anm X,
\label{Ntob}
\eeq
\beq
N_2N_2\to W^+W^-\to \mu\anm,
\label{NtoW}
\eeq
\beq
N_2N_2\to t\bar{t}\to (W^-\to \mu\anm)(\bar{b}\to\mu\anm X),
\label{Ntot}
\eeq
where it is understood that we can have similar channels with the charge conjugate final states of the reactions above. The small branching ratio of the channel of Eq.~(\ref{Ntonu}), as Fig.~(\ref{Br}) shows in the case of the Sun, is partially compensated by its short chain advantage mentioned above.
The distributions of muonic neutrinos for energy $E_{\nm}\equiv E$, are given correspondingly for the channels of interest by
\beq
\frac{dN_{\nm}}{dE}=\delta(E-m) \quad \textrm{for Eq.~(\ref{Ntonu})},
\label{Fnu}
\eeq
\beq
\frac{dN_{\nm}}{dE}=\frac{0.18\cdot 2}{m}\left[1-3\fr{E}{m}{2}+2\fr{E}{m}{3}\right]
\eta(0,m)
\quad \textrm{for Eq.~(\ref{Ntotau})},
\eeq
\beq
\frac{dN_{\nm}}{dE}=\frac{0.13}{\m}\left[\frac{5}{3}-3\fr{E}{\m}{2}+\frac{4}{3}\fr{E}{\m}{3}\right]
\eta(0,\m)
\!, \quad \m=0.58m \quad \textrm{for Eq.~(\ref{Ntoc})},
\label{Fc}
\eeq
\beq
\frac{dN_{\nm}}{dE}=\frac{0.103\cdot 2}{\m}\left[1-3\fr{E}{\m}{2}+2\fr{E}{\m}{3}\right]
\eta(0,\m)
\!, \quad \m=0.73m \quad \textrm{for Eq.~(\ref{Ntob})},
\label{Fb}
\eeq
\beq
\frac{dN_{\nm}}{dE}=\frac{0.107}{m\beta}\eta\left(\frac{m}{2}(1-\beta),\frac{m}{2}(1+\beta)\right)
\!,\quad \beta=\sqrt{1-\frac{m_W^2}{m^2}} \quad \textrm{for Eq.~(\ref{NtoW})}.
\eeq
In the channel of Eq.~(\ref{Ntot}), we have contributions from $W$- and $b$-decays:
\begin{eqnarray}
\lefteqn{ \frac{dN_{\nm}}{dE}=\frac{dN_{\nm(W)}}{dE}+\frac{dN_{\nm(b)}}{dE} \quad \textrm{for Eq.~(\ref{Ntot})}, }\\
& & \frac{dN_{\nm(W)}}{dE}=\frac{0.107}{\left(1-\frac{m_W^2}{m_t^2}\right)m\beta}
\ln\left[\frac{\min\left(\frac{Em_t}{m(1-\beta)},\frac{m_t}{2}\right)}
{\max\left(\frac{Em_t}{m(1+\beta)},\frac{m_W^2}{2m_t}\right)}\right]\times \nonumber\\
& & \qquad \times\eta\left(\frac{mm_W^2}{2m_t^2}(1-\beta),\frac{m}{2}(1+\beta)\right),\\
\label{Ftb}
& & \frac{dN_{\nm(b)}}{dE}=\frac{0.103\cdot 2}{\left(1-\frac{m_W^2}{m_t^2}\right)\m\beta}
\left[F(E,E_-,E_+)\eta(0,E_-)+F(E,E,E_+)\eta(E_-,E_+)\right],\\
& & F(E,E_1,E_2)=\frac{2}{3}\left[\fr{E}{E_1}{3}-\fr{E}{E_2}{3}\right]-\frac{3}{2}\left[\fr{E}{E_1}{2}-\fr{E}{E_2}{2} \right]+\ln\frac{E_2}{E_1}, \nonumber\\
& & \m=0.73m, \quad E_{\pm}=\frac{m_t^2-m_W^2}{2m_t}\m(1\pm\beta), \quad \beta=\sqrt{1-\frac{m_t^2}{m^2}}.\nonumber
\end{eqnarray}
The step function is
$$\eta(E_1,E_2)\equiv \left\{
\begin{array}{ll}
1 & \textrm{for }E_1<E<E_2,\\
0 & \textrm{otherwise.}
\end{array} \right.$$
Note, that most of the formulas are taken from \cite{Jungman,Hooper}. The values $\bar{m}<m$ in Eqs.~(\ref{Fc},~\ref{Fb},~\ref{Ftb}) take into account the partial energy losses by $c$- and $b$-quark respectively, while they are hadronized. For the $WW$-channel, the effect of $W$-polarization was neglected. Consideration of this effect would correct our estimation for $\nm$ from this channel by $\sim 20$\%. However, as we shall see, it is  not necessary to consider this correction because for $m\gsim 3$ TeV, where the $WW$-channel is important, the predicted muon signal becomes too faint for the existing experimental setups. Also note, that the spectrum of Eq.~(\ref{Ftb}) for $\nm$ from $b$-decay in Eq.~(\ref{Ntot}) differs from that given in \cite{Jungman,Hooper}. In the notation of \cite{Jungman,Hooper}, the signs before $x^2$ and $y^2$ in the respective formulas there should be altered in order to be correct.
All spectra predicted by Eqs.~(\ref{Fnu}--\ref{Ftb}), are illustrated in Fig.~(\ref{spectra}).
The effect of absorbtion of $\nm$ in the solar matter is taken into account following \cite{Crotty}
\begin{eqnarray}
\frac{dN_{\nm}(\textrm{outside Sun})}{dE}=\frac{dN_{\nm}(\textrm{in the Solar core})}{dE}\exp\left(-\frac{E}{130\GeV}\right),\\
\frac{dN_{\anm}(\textrm{outside Sun})}{dE}=\frac{dN_{\anm}(\textrm{in the Solar core})}{dE}\exp\left(-\frac{E}{200\GeV}\right).
\end{eqnarray}
Note, that the spectra in the solar core for neutrinos $\nm$ and antineutrinos $\anm$ do not differ and are given by Eqs.~(\ref{Fnu}--\ref{Ftb}).

\begin{figure}
\begin{center}
\includegraphics[scale=0.6]{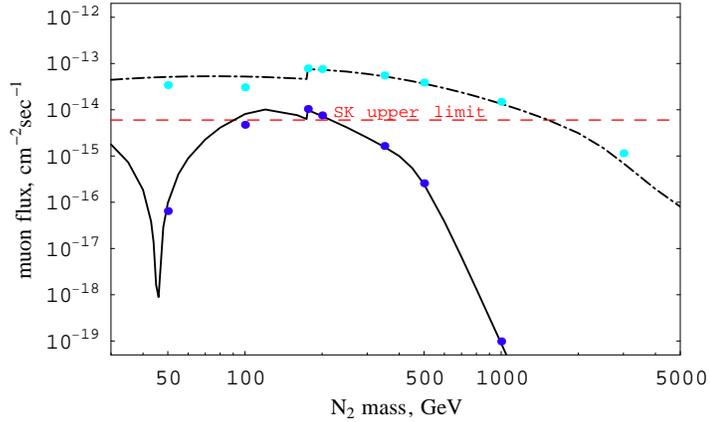}
\caption{Muon fluxes predicted for the models \cite{CK} (solid line) and \cite{fin} (dot-dashed line) in comparison to the Super-Kamiokande constraint. Dark and light dots show predictions based on the results of \cite{Edsjo} for the models \cite{CK} and \cite{fin} respectively.}
\label{Fmuon}
\end{center}
\end{figure}

The muon (including both $\mu^-$ and $\mu^+$) fluxes predicted for the models \cite{CK} and \cite{fin} are shown in Fig.~(\ref{Fmuon}). We compare them with the respective upper limit obtained by Super-Kamiokande (SK) \cite{SK}
$$\Phi_{\mu}<6\cdot 10^{-15} \cm^{-2}\s^{-1}.$$
It relates to an angle of 10$^\circ$ around the Solar center, that is expected to embrace most of the muon flux induced by annihilation in the Sun for the \N2 masses of interest.
For our chosen parameters, the SK limit excludes the intervals
\beq
100\GeV<m<200\GeV
\label{mexcl-1}
\eeq
for the model \cite{CK} and
\beq
m<1500\GeV
\label{mexcl-2}
\eeq
for the model \cite{fin}.

\section{Discussion}
In the present paper we have considered candidates emerging
from the Minimal Walking Technicolor model with a suppressed
coupling to the $Z$-boson. These candidates can account for the dark matter
density of the Universe and simultaneously can avoid any contradiction
with the results of direct dark matter search experiments. In fact CDMS and Xenon experiments can exclude
only a tiny window around 100 GeV for the model \cite{CK}.
Being elusive for direct dark matter
searches, we investigated the possibility of indirect effects of $N_2$ dark matter particles. In particular,
we estimated the neutrino fluxes on the surface of the Earth from $N_2$ annihilations in the Sun and in the Earth. These effects can provide constraints on the parameters of the considered
models.

One of the biggest uncertainties entering these constraints is that the final
results depend on the local density of DM particles as well as
on the velocity distribution. As one can see in Fig.~(\ref{Fmuon}), a decrease of
$\rho_{loc}$ down to 0.2 GeV/cm$^3$, a value that is currently acceptable, with unchanged velocity distribution, would leave
only a tiny interval of $m$ around 110 GeV excluded for the model
\cite{CK}. It makes indefinite any conclusion for the model \cite{CK},
based on the searches for muon signals. In any case, the obtained constraint is more strict that the one
based on direct dark matter search experiments~\cite{Kouvaris:2008hc}.

Regarding the model \cite{fin}, the existing uncertainties can
hardly influence essentially the excluded range of Eq.~(\ref{mexcl-2}).
This result agrees with the analogous result, obtained
by the Kamiokande collaboration for Majorana neutrino \cite{Kam},
although it differs in some details (which are most likely related to
the estimations of cross sections for different annihilation
channels and their neutrino yields). Moreover, the effect of $b$- and
$c$-quarks slowing down in dense Solar matter, neglected in our
consideration, and all the unaccounted annihilation chains, producing
$\nm$ including chains which go through $b$- and $c$-quarks, can
lead to an increase of the predicted muon fluxes for large $m$.
 This can give neutrinos with lower energy, which may avoid absorbtion in the Solar matter.
For this reason the channel $N_2N_2\to
\bar{t}t \to b\to \nm X$, giving softer $\nm$ than $\nm$ from $t\to W$-
and $WW$-channels (see Fig.~(\ref{spectra})), has a significant
contribution for high $m$.

Uncertainties come also from neutrino propagation effects, i.e. neutrino oscillations
 and interactions with matter. Oscillation effects in vacuum and matter \cite{MSW} might change
  the flavor content of the neutrino flux coming from the Solar core to the detectors on the Earth.
The uncertainties in the description of oscillations become less ambiguous once we average the effect because of the large distance, energy and time measurement intervals involved.
Indeed, neutrinos of three flavors are generated in the source. The flux of muon neutrinos at the Earth is
\beq
\Phi_{\nm}=P_{\mu\mu}\Phi_{\nm}+P_{e\mu}\Phi_{\nu_e}+P_{\tau\mu}\Phi_{\nu_{\tau}},
\eeq
where $P_{\alpha\beta}$ is the probability of transition between flavors $\alpha$ and $\beta$.
As we see in the right pannel of Fig.~(\ref{Br}), the most important \N2 annihilation channels of neutrino production, depending on the mass, are $N_2N_2\to \bar{b}b$ with $b$ decaying into $cl\bar{\nu}$, $N_2N_2\to \bar{t}t$ with $t\to Wb$, and $N_2N_2\to WW$ with $W\to l\bar{\nu}$.
In order to give a simple estimate of the effect of neutrino oscillations, we are going to consider the $\bar{b}b$-mode as the dominant one for $m<200$ GeV, the $\bar{t}t$-mode for $0.2<m<3$ TeV, and the $WW$-mode for $m>3$ TeV.
In the $\bar{b}b$-mode, the production of $\nu_{\tau}$ is suppressed because of the fact that $\tau$ is much heavier
than the muon and the electron (we ignore here the difference in the neutrino spectra, and we also ignore the production of
$\nu_{\tau}$ from $\tau$-decay), so the produced neutrino flux is roughly
$\nu_e+\nm+0.3\nu_{\tau}$. In the $WW$-mode, one has $\nu_e+\nm+\nu_{\tau}$. Finally in the $\bar{t}t$-mode both $W$ and $b$ decay, so we have roughly $\nu_e+\nm+0.7\nu_{\tau}$.
For the transition probabilities obtained in \cite{Cirelli} within the 3-flavor scheme we have:
all $P_{\alpha\beta}\approx 0.3$ for $E\lsim 10$ GeV, and
$P_{\mu\mu}=P_{\tau\mu}=P_{\tau\tau}\approx 0.4$,
$P_{\tau e}=P_{\mu e} \approx 0.2$ and $P_{ee}\approx 0.6$ for $E\gg 10$ GeV. Therefore
the following change of the flavor content of neutrino flux coming from the Solar core to the detector might take place
$$\nu_e+\nm+0.3\nu_{\tau}\Longrightarrow \left\{
\begin{array}{ll}
0.8\nu_e+0.8\nm+0.8\nu_{\tau},  & \textrm{at }E\lsim 10 \GeV\\
0.9\nu_e+0.7\nm+0.7\nu_{\tau},  & \textrm{at }E\gg 10 \GeV
\end{array} \right.\textrm{for }m\lsim 200\GeV,$$
$$\nu_e+\nm+0.7\nu_{\tau}\Longrightarrow
0.9\nu_e+0.9\nm+0.9\nu_{\tau},  \,\textrm{at any }E,
\textrm{ for }200\GeV\lsim m\lsim 3\TeV,$$
$$\nu_e+\nm+\nu_{\tau}\Longrightarrow
\nu_e+\nm+\nu_{\tau}, \,\textrm{at any }E,
\textrm{ for }m\gsim 3\TeV.$$
In our rough estimate we assume that $P_{\alpha\beta}=P_{\beta\alpha}=(P_{\alpha\beta}+P_{\bar{\alpha}\bar{\beta}})/2$.
In this simplified picture we see that oscillations re-distribute the flavor content of the neutrino flux, making it more homogenous. This would decrease the predicted $\nm$-flux by $\sim 20$\%.

Effects of $\nu$ interactions lead to not only absorbtion of neutrinos in the Solar matter due to charge current (CC) interactions, but also to loss of energy for $\nu_{e,\mu}$ due to neutral current (NC) interactions, and for $\nu_{\tau}$ due to both NC and CC (in the latter case $\nu_{\tau}$ is re-generated from the chain $\nu_{\tau}N\to \tau X$, $\tau \to \nu_{\tau}X$). Energy loss is a small effect (the ratio of respective cross sections is $\sigma_{NC}/\sigma_{CC}\sim 1/3$). It should decrease a little the muon signal for small $m$, but increase it a bit for large $m$. This is analogous to the neutrinos produced from long cascade chains that we neglected, since a shift to a lower energy saves partially neutrinos from absorption.

In the 4-flavor oscillation scheme, where a sterile type of neutrino $\nu_s$ is added, there can be also a damping effect of the signal for low \N2 mass, because of the $\nm\to \nu_s$ transition. However an amplification of the signal for high mass can occur because of the larger penetrating ability of neutrinos oscillating to $\nu_s$ \cite{Naumov}.

In \cite{Edsjo} (and the therein referred web-site), the muon flux for separate annihilation channels are
obtained taking into account oscillation and interaction effects. For the sake of comparison,
we have taken their data on muon fluxes for the case of ``standard" oscillation parameters (case ``B" in their notation).
For a \N2 mass $m<175$ GeV we considered the channels $b\bar{b}$, $c\bar{c}$, $\tau\tau$, for $175<m<2000$ GeV the $t\bar{t}$ channel, and
for $m>2000$ GeV the $t\bar{t}$ and $WW$ ones. Since the data were related to muon fluxes per one annihilation act, we multiplied the fluxes by the respective annihilation rates. The points on Fig.~(\ref{Fmuon}) show the respective results for a few mass values. As seen, the agreement is extremely good especially for high \N2 mass. Therefore the upper limits on $m$ in Eqs.~(\ref{mexcl-1},~\ref{mexcl-2}) do not appreciably change. At lower mass, (which is of limited interest for the studied models), where the $b\bar{b}$-channel dominates, a difference slightly larger is noted between our result and~\cite{Edsjo}, mainly due to ignorance of the $b$-quark energy losses in our calculation and the aforementioned oscillation effect. However, the suppression of $\nm$-yields from the $b\bar{b}$-channel is partially compensated by the excessive $\nm$-yields from the $\tau\tau$-channel scaling as 4\% of the branching ratio at this particular \N2 mass range. 


\section*{Acknowledgements}

The work of KB was supported by the grants of Khalatnikov-Starobinsky leading scientific school N 4899.2008.2, and of Russian leading scientific school N 3489.2008.2. The work of CK was supported by the Marie Curie Fellowship under contract MEIF-CT-2006-039211. We also would like to thank J.~Edsj\"o for helping with the interpretation of the results published in~\cite{Edsjo}.

\end{document}